# Electronic Structure and Transport in the Potential Luttinger Liquids CsNb$_3$Br$_7$S and RbNb$_3$Br$_7$S

Fabian Grahlow[a], Fabian Strauß[b], Marcus Scheele[b], Markus Ströbele[a], Alberto Carta[c], Sophie F. Weber[c], Scott Kroeker[d], Carl P. Romao*[c] and H.-Jürgen Meyer*[a]

The crystal structures of ANb$_3$Br$_7$S (A = Rb and Cs) have been refined by single crystal X-ray diffraction, and are found to form highly anisotropic materials based on chains of the triangular Nb$_3$ cluster core. The Nb$_3$ cluster core contains seven valence electrons, six of them being assigned to Nb-Nb bonds within the Nb$_3$ triangle and one unpaired d electron. The presence of this surplus electron gives rise to the formation of correlated electronic states. The connectivity in the structures is represented by one-dimensional [Nb$_3$Br$_7$S]$^-$ chains, containing a sulphur atom capping one face ($\mu_3$) of the triangular niobium cluster, which is believed to induce an important electronic feature. Several types of studies are undertaken to obtain deeper insight into the understanding of this unusual type of material: the crystal structure, morphology and elastic properties are analysed, as well the (photo-) electrical properties and NMR relaxation. Electronic structure (DFT) calculations are performed in order to understand the electronic structure and transport in these compounds, and, based on the experimental and theoretical results, we propose that the electronic interactions along the Nb chains are sufficiently one-dimensional to give rise to Luttinger liquid (rather than Fermi liquid) behaviour of the metallic electrons.

## Introduction

Metal-rich niobium halides are well known in chemistry, typically appearing with the octahedral Nb$_6$ cluster core, as in Nb$_6$X$_{14}$ (X = Cl, Br),[1,2,3] or with a triangular Nb$_3$ cluster core in the structure of Nb$_3$X$_8$ (X = Cl, Br, I).[4,5] The electronic structure of Nb$_6$X$_{14}$ is well described with eight metal-centred molecular orbitals (MOs) being occupied by 16 electrons. Nb$_3$X$_8$ (X = Cl, Br, I) compounds appear as layered structures based on close-packed halide arrangements in which ¾ of the octahedral voids in every second interlayer are occupied by niobium atoms. However, each niobium atom in the crystal structure is displaced away from the centre of its coordination octahedron to form trigonal clusters (shown in Figure 1), each containing seven electrons in metal-centred states.[6] A magnetic-to-nonmagnetic phase transition has been shown upon cooling in Nb$_3$Cl$_8$ with interlayer charge transfer between Nb$_3$ trimers in adjacent layers,[7] although such charge transfer was not observed in other studies.[8,9]

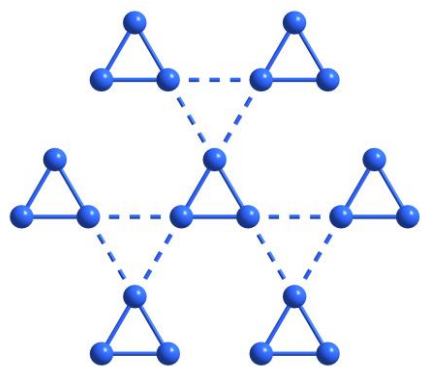

Figure 1: Projection of arrangement of Nb$_3$ clusters within one interlayer of the structure of Nb$_3$X$_8$.

The valence electron count (VEC) in compounds containing triangular Nb$_3$ clusters compounds is between 6 and 8; for instance, VEC = 6 for Nb$_3$Br$_7$S[10] and Nb$_3$O$_2$Cl$_5$[11], VEC = 7 for Nb$_3$X$_8$ (X = Cl, Br, I) and CsNb$_3$Br$_7$S[12], and VEC = 8 for NaNb$_3$Cl$_8$.[13,14,9] The compounds with seven valence electrons can have interesting magnetic behaviour, especially in combination with the two-dimensional polar Kagome lattice structure of Nb$_3$X$_8$.[8,9,15-17] MO calculations on the seven-electron cluster [Nb$_3$X$_{13}$]$^{5-}$, adapted from the Nb$_3$X$_8$ structure, revealed that six electrons can be assigned to Nb-Nb bonding within the Nb$_3$ cluster and that the HOMO level is a half-filled 2a$_1$ orbital.[6]

[a.] Section for Solid State and Theoretical Inorganic Chemistry
Institute of Inorganic Chemistry
Eberhard-Karls-Universität Tübingen
Auf der Morgenstelle 18, 72076 Tübingen, Germany
*[a.] E-mail: juergen.meyer@uni-tuebingen.de
[b.] Institute for Physical and Theoretical Chemistry
Eberhard-Karls-Universität Tübingen
Auf der Morgenstelle 18, 72076 Tübingen, Germany
[c.] Department of Materials, ETH Zurich,
Wolfgang-Pauli-Str. 27, 8093 Zürich, Switzerland
*[c.] E-mail: carl.romao@mat.ethz.ch
[d.] Department of Chemistry, University of Manitoba, Winnipeg, Manitoba R3T 2N2, Canada

Compounds A$_3$[Nb$_6$SBr$_{17}$] with A= Rb, Tl, K, Cs[18, 19] contain a sulphur-centred, trigonal prismatic [Nb$_6$S] core which has been reported to contain only weak Nb-Nb interactions between adjacent Nb$_3$ triangles (Figure 2 at left).[18] The chain structure of CsNb$_3$SBr$_7$ is based on similar [Nb$_3$S] clusters, shown in Figure 2 on the right. Both compounds contain seven valence electrons per Nb$_3$ cluster.

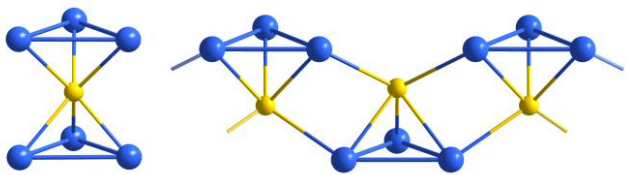

Figure 2: The [Nb$_6$S] cluster core in the structure of A$_3$[Nb$_6$SBr$_{17}$] (left), and a section of the infinite [Nb$_3$S] chain in ANb$_3$Br$_7$S (right).

Triangular [Nb$_3$]$^{8+}$ clusters consisting of an Nb-Nb bonded arrangement are of fundamental interest. In isolation, each cluster has seven valence electrons and thus one unpaired d electron in a 2a$_1$ orbital. Such unpaired electrons in Nb clusters have recently become the source of significant research attention due to their formation of various long-range ordered states, for example by charge disproportionation between layers in Nb$_3$Cl$_8$[7], and the formation of a Mott insulating state in GaNb$_4$S$_8$.[20]

The structure of CsNb$_3$Br$_7$S was reported in 1993 to contain [Nb$_3$]$^{8+}$ clusters, each of which is capped with a ($\mu_3$) sulphur atom to form the motif of a tetrahedron (Figure 2, right). These clusters are arranged in chains parallel to the *c*-axis, with alternating positions of the clusters above and below the central sulphur chain. The chains have an outer shell of bromide ions, and voids in the crystal structure between the chains are filled by caesium ions. The extended Hückel method was used to predict that CsNb$_3$Br$_7$S is a semimetal, with the Nb d orbitals forming bands which cross the Fermi level.[12]

We have revisited this material, which shows an interesting crystal morphology of cleavable rods. The synthesis and crystal structure of CsNb$_3$Br$_7$S is revised in order to confirm its structure and electronic properties. In addition, the synthesis and structure of the closely related compound RbNb$_3$Br$_7$S is reported for the first time. The elastic properties and NMR relaxation of CsNb$_3$Br$_7$S are investigated, electric conductivity measurements along the [Nb$_3$S]$_n$ chain direction are performed for both compounds, and theoretical investigations of the electronic structure are performed using density functional theory (DFT).

Our studies suggest that CsNb$_3$Br$_7$S and RbNb$_3$Br$_7$S have conducting electrons which can be described as (Tomonaga–) Luttinger liquids, rather than the more usual Fermi liquids. Luttinger liquids are a type of one-dimensional paramagnetic quantum fluid, characterized by their separate transport of charge density waves (CDWs) and spin density waves (SDWs).[21,] [22] They can be identified by their characteristic spectral function and by the power-law response of various physical properties as a function of temperature.[21]

## Experimental

### Synthetic procedures

ANb$_3$Br$_7$S compounds with A = Rb and Cs were prepared by solid-state reactions starting from mixtures of ABr (99.0% Merck), NbBr$_5$ (99.9% ABCR GmbH), niobium powder (99.9% ABCR GmbH) and sulphur (99.9% Carl Roth GmbH) in stoichiometric proportions with total masses of 100 to 200 mg per batch. The starting materials were homogenised by grinding under an argon atmosphere and fused in silica tubing. Loaded ampoules were heated at 800 °C for 24 h (heating and cooling rates of 2 K/min). Short-term reactions and fast cooling rates yielded a furry, crystalline material consisting of fine needle-like crystals. Long-term reactions with slow cooling rates gave longer, rod-shaped crystals. Both compounds ANb$_3$Br$_7$S (A = Rb, Cs) appear as black crystals with metallic lustre and are stable in air for weeks, as confirmed by X-ray powder diffraction.

### Crystallography

Carefully selected black needle-shaped single-crystals of RbNb$_3$Br$_7$S and CsNb$_3$Br$_7$S were mounted on a Rigaku XtaLab Synergy-S X-ray diffractometer using Cu-K$\alpha$ ($\lambda$ = 1.54184 Å) and Mo-K$\alpha$ ($\lambda$ = 0.71073 Å) radiation, respectively. The single crystals were kept under N$_2$ cooling at 100 K during the data collection. Corrections for absorption effects were applied with CrysAlisPro 171.41.64.93a (Rigaku Oxford Diffraction, 2020). Crystal structures were solved by the integrated space group and crystal-structure determination routine of SHELXT[23] and full-matrix least-squares structure refinements with SHELXL-2014[23] implemented in Olex2 1.3-ac4.[24] The structure of RbNb$_3$Br$_7$S was refined as a racemic twin with a Flack parameter 0.50(3).

Details of the crystal structure investigations can be obtained from the joint CCDC/FIZ Karlsruhe online deposition service: https://www.ccdc.cam.ac.uk/structures/ by quoting the deposition numbers CSD-2048759 for RbNb$_3$Br$_7$S and CSD-2048757 for CsNb$_3$Br$_7$S.

### Electron Microscopy

Electron micrographs were recorded on a JEOL 8900 Superprobe spectrometer. The samples were coated with carbon prior to the measurements.

### Electrical Characterisation

Conductivity measurements were performed in a Lake Shore Cryotronics CRX-6.5K probe station with a Keithley 2636B source meter unit. Rod-shaped crystals of CsNb$_3$Br$_7$S and RbNb$_3$Br$_7$S were transferred into the chamber under protective gas and contacted with silver paste on a silicon substrate with 770 nm oxide layer. The conductive silver pads at each end of the crystals were connected to the circuit with tungsten tips (Figure S1). The chamber was kept under vacuum (<5·10$^{-5}$ mbar) and the temperature was varied between 20 K and 300 K. Before each measurement, sufficient time was allowed for the sample to reach the chosen temperature. Two-point conductivity measurements were performed by varying the applied source-drain voltage from -1 V to 1 V while detecting the current. Time-resolved photocurrent measurements used a picosecond pulsed laser driver (Taiko PDL M1, PicoQuant) together with a laser head 779 nm (pulse length <500 ps); the crystals were illuminated at 40 mW laser output power using the continuous wave mode under a constant bias of 1 V.

**Nuclear Magnetic Resonance Spectroscopy**

$^{133}$Cs NMR data were collected at 65.5 MHz ($B_0$ = 11.7 T) and 52.4 MHz ($B_0$ = 9.4 T) using Bruker Avance III spectrometers. The $^{133}$Cs magic-angle spinning (MAS) spectra were acquired on a polycrystalline sample at 11.7 T by Bloch decay, using a 2.5 mm MAS probe with spinning rates of 10.00, 20.00 and 30.00 kHz. The chemical shift is referenced to external 0.1 M CsCl(aq) at 0.0 ppm. $T_1$ relaxation times were measured by fitting inversion-recovery data to a mono-exponential decay function.

$^{93}$Nb NMR spectra were acquired at 97.9 MHz ($B_0$ = 9.4 T) on a Bruker Avance III 400 using a 4 mm MAS probe. The wideband uniform-rate smooth-truncation (WURST) quadrupolar Carr-Purcell-Meiboom-Gill (QCPMG) pulse sequence was used to acquire six individual spectra on a non-spinning sample at variable transmitter offsets, which were subsequently assembled into the 1.8 MHz spectrum. All NMR spectra were acquired at ambient temperature without temperature regulation.

**Computational Methods**

Density functional theory (DFT) calculations were performed using the software packages Abinit (v. 9)[25] and Quantum Espresso. Calculations of the elastic tensor and Γ-point phonon frequencies were performed in Abinit using norm-conserving pseudopotentials from the Abinit library, a 4 × 4 × 4 Monkhorst–Pack grid of **k**-points,[26] a 34 Ha plane-wave basis set energy cutoff, and the PBE exchange–correlation functional.[27]

Calculations of the electronic band structure with antiferromagnetic ordering and a Hubbard U term[28] (U = 5 eV and J = 0.2 eV on the Nb sites)[29] and with spin–orbit coupling were performed in Abinit using the projector-augmented wave (PAW) method[30] with pseudopotentials from the GBRV library,[31] a Monkhorst–Pack grid of **k**-points with real-space basis vectors [0 2 4] [4 0 4] and [4 2 0],[26] a 128 Ha plane-wave basis set energy cutoff within the PAW spheres and a 24 Ha cutoff outside.

We also performed calculations without spin polarization to construct a tight-binding model based on Wannier functions. For these calculations we employed Wannier90[32] and Quantum Espresso[33] using ultrasoft pseudopotentials also from the GBRV library [33] with a 4 × 2 × 6 Monkhorst–Pack grid, and a plane-wave energy and density cutoff of 72 Ry and 864 Ry, respectively.

All the calculations stated above were performed with PBE exchange–correlation functional [27] with the DFT-D3 dispersion correction,[34] and Methfessel–Paxton cold smearing of the electronic states.[35] Special points in and paths through the Brillouin zone were chosen following Hinuma et al.[36] All computational parameters were chosen following convergence studies. The band structures produced by Abinit and by Quantum Espresso with these parameters were compared and found to be essentially identical.

## Results and Discussion

**Crystal Structures**

The crystal structures of ANb$_3$Br$_7$S (A = Rb, Cs) were solved and refined from single-crystal X-ray diffraction data. The structure of CsNb$_3$Br$_7$S was confirmed to agree with the previously reported data on the same compound with the monoclinic space group $P2_1/c$ (No. 14).[12] The structure of the new compound RbNb$_3$Br$_7$S was refined with the orthorhombic space group $Pmc2_1$ (No. 26). Both crystal structures contain the same arrangement of triangular Nb$_3$ clusters face-capped (μ$_3$) by sulphur atoms, and linked into chains via bromide and sulphide ions along the crystallographic *a* direction for A = Rb, and along *c* direction for A = Cs. The [Nb$_3$SBr$_7$]$^-$ cluster units in both structures are linked into chains by three inner (i) edge capping bromides, six shared (6/2) outer (a-a) bromides, and one terminal outer (a) bromide, forming the structural unit [Nb$_3$SBr$_3^i$Br$_{6/2}^{a-a}$Br$^a$]$^-$, which can be seen in Figure 3.

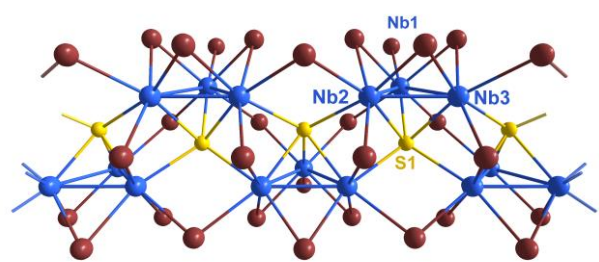

Figure 3: Section of a single chain in the structure of CsNb$_3$Br$_7$S along the crystallographic *c*-direction. Niobium atoms are shown blue, sulphur yellow, and bromide atoms brown.

The difference between the two crystal structures with A = Cs and Rb is expressed in the different symmetries of the orthorhombic and monoclinic space groups. This can be explained by the different radii and coordination patterns of alkali ions in the structures. The caesium ion is situated in a nearly regular cube-octahedral environment of bromides, and Rb is surrounded by an irregular twelvefold arrangement of bromide ions. These environments with bromide ions, which are also linked to the niobium clusters, cause shifts of adjacent clusters along the one-dimensional chain of the structure (Figure 5).

The distances between the three crystallographically distinct niobium atoms are about 290 pm within $Nb_3$ cluster triangles and about 310 pm between adjacent triangles (see Table 1 for details). Despite the similarity of these interatomic distances, previously reported electronic structure calculations (extended Hückel) on $CsNb_3Br_7S$ had revealed comparable Nb-Nb crystal orbital overlap populations (0.25) within $Nb_3$ triangles and weak (0.09) overlap populations between adjacent triangles.

Table 1. Selected interatomic distances (pm) in $ANb_3Br_7S$ (A = Rb, Cs) compounds. The A = Rb compound contains two independent $Nb_3$ clusters in the structure.

a) within cluster triangle ($\Delta$)
b) between cluster triangles (')

| Compound | $RbNb_3Br_7S$ | $CsNb_3Br_7S$ |
| --- | --- | --- |
| Nb-Nb $\Delta$ a) | 289.3(3), 290.3(3) | 290.71(7) |
|  | 289.3(3), 290.3(3) | 290.01(7) |
|  | 281.2(3) 288.6(3) | 286.63(7) |
| Nb-Nb ' b) | 310.2(3) | 310.74(5) |
| Nb-S $\Delta$ | 238.8(6) 248.4(8) | 238.3(2) |
|  | 238.8(6) 248.4(8) | 246.0(2) |
|  | 233.0(8) 250.7(9) | 246.1(2) |
| Nb-S ' | 257.2(6) | 262.9(2) |
|  | 265.6(5) | 263.5(2) |

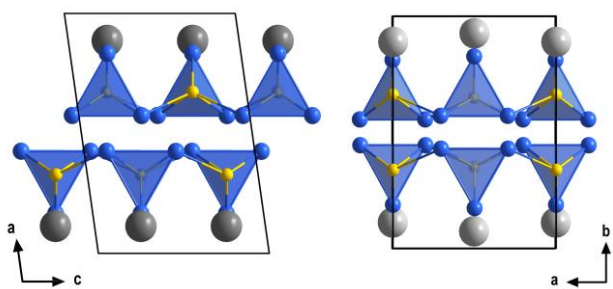

Figure 5: Comparison of crystal structures of $CsNb_3Br_7S$ ($P2_1/c$, left) and $RbNb_3Br_7S$ ($Pmc2_1$, right) with their unit cells. Bromides are omitted in the drawings for better clarity. Cs (dark grey) and Rb (light grey) atoms are shown.

The cohesion between adjacent cluster chains in the structure is dominated by ionic (Br-A) bonding with the A cations, as expressed by the appearance of differently shifted chain arrangements in the crystal structures of $ANb_3Br_7S$ with A = Rb and Cs along the chain directions (Figure 5). The primary growth direction of needle-shaped crystal rods coincides with the direction of the $[Nb_3SBr_3^iBr_{6/2}^{a-a}Br^a]^-$ chains (Figure 4). This pronounced growth direction suggests a one-dimensional character of the material, as supported by the fraying behaviour of crystal rods shown below, and later quantitatively evaluated by the elastic properties (directional Young's modulus).

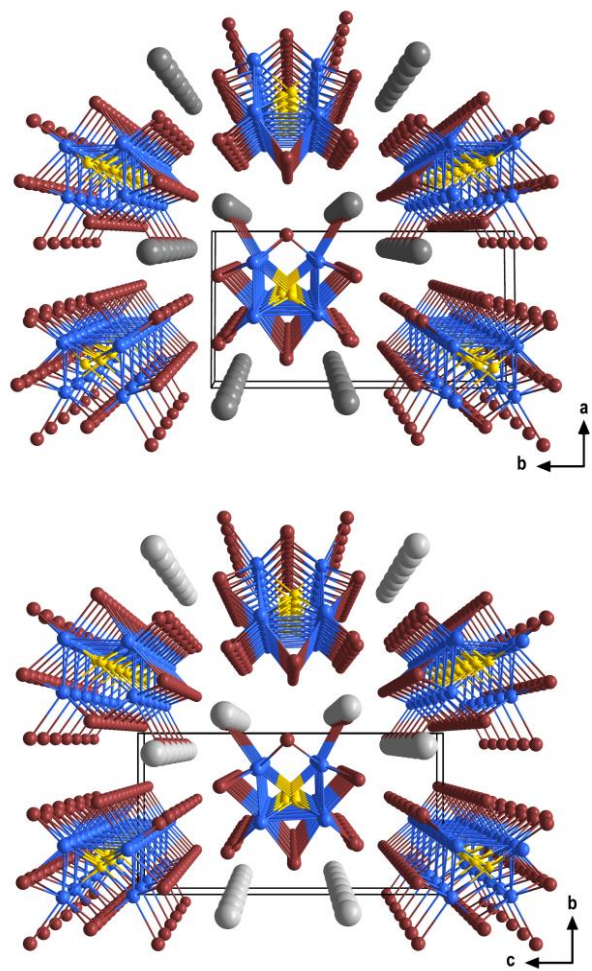

Figure 4: Comparison of crystal structures of $RbNb_3Br_7S$ ($Pmc2_1$, bottom) and $CsNb_3Br_7S$ ($P2_1/c$, top) with their unit cells. Cs (dark grey) and Rb (light grey) atoms are shown.

**Electron Microscopy and Elastic Properties**

The morphology of synthesized $CsNb_3Br_7S$ crystals is shown in SEM micrographs in Figure 6. $CsNb_3Br_7S$ forms rods with fairly uniform dimensions of about 500 µm length and 10 µm diameter. As Figure 6a shows, each rod could be assumed to be a single crystal, but a magnified view suggests that this is not always clear. This morphology can be readily explained by a highly anisotropic crystal structure, and our XRD studies on single crystals confirm the long axis of the rods to coincide with the crystallographic *c*-axis. Figure 6b shows fraying of a rod, an interesting feature indicating that the material is at least somewhat flexible, and that the cohesive forces along the a and b axes might be expected to have van der Waals character; except in this case they are instead ionic interactions between the bromides of the cluster and A cations.

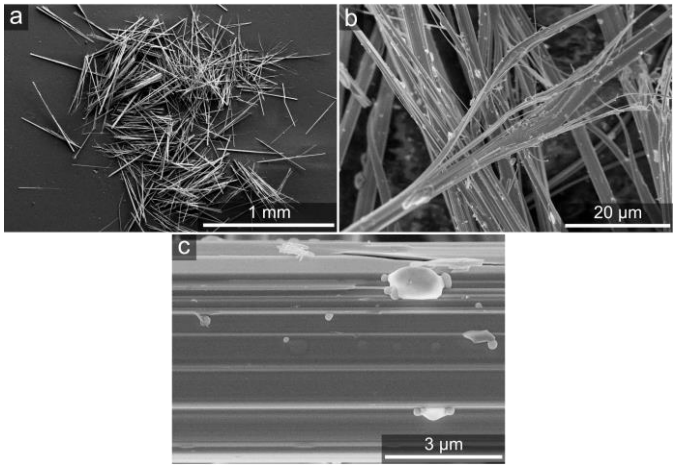

Figure 6: SEM micrographs of (a) $CsNb_3Br_7S$ crystals, showing their pronounced rod-like morphology, (b) fraying of a rod of $CsNb_3Br_7S$ and (c) their high degree of uniformity on the micron scale.

In order to examine the origins of this fraying behaviour further, and to determine to what extent the elastic properties of $CsNb_3Br_7S$ can be considered one-dimensional, its elastic tensor was calculated using DFPT (Eq. S1). The vdw-D3 dispersion correction of Grimme was employed to ensure that van der Waals forces would be accounted for in the model. The easiest way to visualize the elastic anisotropy is through the directional Young's modulus ($Y_{ii}$), which describes the resistance of the material to uniaxial stress in a given direction. This is shown in Figure 7; $CsNb_3Br_7S$ is found to be stiffest along the c-axis, with a directional Young's modulus of 74 GPa. While c is the stiffest direction, the Young's modulus also shows significant peaks coinciding with a (24 GPa) and b (34 GPa), indicating that the mechanical bonding between $[Nb_3Br_7S]^-$ chains through the $Cs^+$ cations is significant. These results demonstrate that, although $CsNb_3Br_7S$ is highly anisotropic, its elasticity is decidedly not one-dimensional.

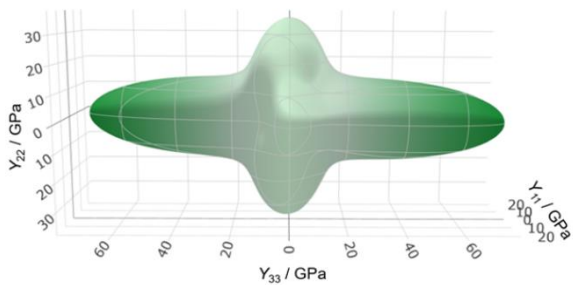

Figure 7: The calculated directional Young's modulus ($Y_{ii}$) of $CsNb_3Br_7S$, shown as a green surface. The view is along the a-axis (analogous to Figure 2).

### Electrical Properties

Both $CsNb_3Br_7S$ and $RbNb_3Br_7S$ exhibit ohmic behaviour in the measured temperature range between 20 and 300 K. Figure 8 shows the dark current obtained at room temperature for two different $CsNb_3Br_7S$ crystals. Conductivities of 0.35 S/cm for $CsNb_3Br_7S$ and 1.65 S/cm for $RbNb_3Br_7S$ are measured with a two-point measurement. Figure 9 shows the temperature-dependent electrical conductivity in the range of 20 K to 300 K for $CsNb_3Br_7S$, indicating a temperature-activated behaviour. The same trend of a decreasing conductivity with decreasing temperature is observed for $RbNb_3Br_7S$, cf. Figure S2.

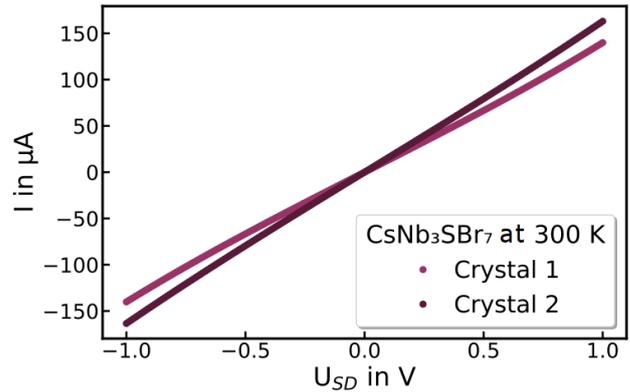

Figure 8: Dark currents I of $CsNb_3Br_7S$ crystals on silicon with 770 nm dioxide layer at 300 K with an applied source-drain Voltage $U_{SD}$ of -1 V to 1 V.

The electrical conductivity of $CsNb_3Br_7S$ (Figure 9) and $RbNb_3Br_7S$ (Figure S2, right) show a general increase with increasing temperature. As shown in Figure 10, above 50 K ($CsNb_3Br_7S$) and 30 K ($RbNb_3Br_7S$), the electrical conductivity ($\sigma$) as a function of temperature can be fitted to a power law ($\sigma(T) = cT^\alpha$ for some constants $\alpha$ and $c$), and therefore is consistent with a Luttinger liquid.[21, 37] At lower temperatures, the conductivity cannot be fit to an Arrhenius curve, suggesting that it is not a simple semiconductor in that regime.

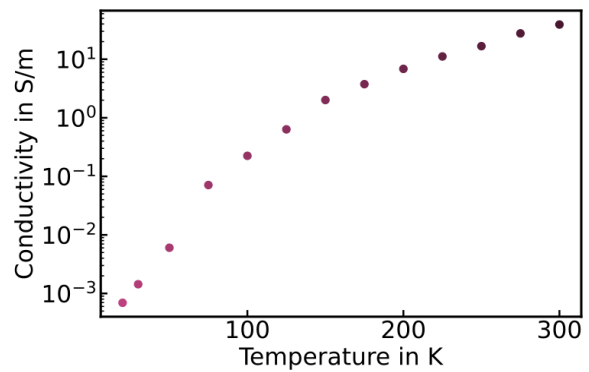

Figure 9: Electrical conductivity of $CsNb_3Br_7S$ versus temperature in a range of 20 K to 300 K.

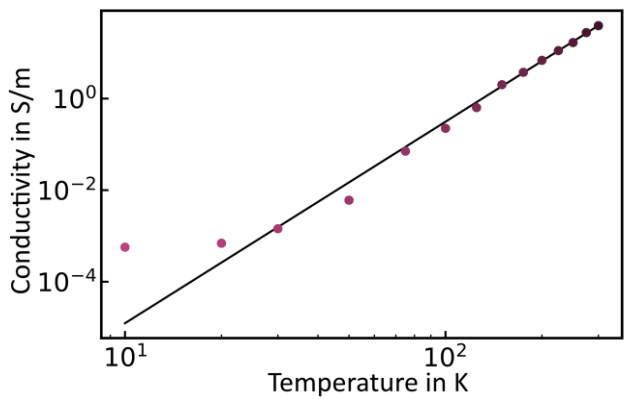

Figure 10: Double logarithmic plot of the electrical conductivity of CsNb$_3$Br$_7$S versus set temperature in a range of 20 K to 300 K with a power law fit (black) following $\sigma(T) = cT^\alpha$ with $\alpha = 4.4$.

**Photoelectric Properties**

When the crystals are illuminated with 779 nm CW-laser pulses, both compounds show a photoresponse, see Figure 11. This photocurrent consists of a steep and fast increase, followed by a slower second component. We attribute the first rise to photoexcited charge carriers, and the second increase to the effect of concomitant heating of the crystal (compare with Figure 8). The photocurrent is strongly temperature-activated, ranging from a few nA at 20 K to several tens of µA at room temperature.

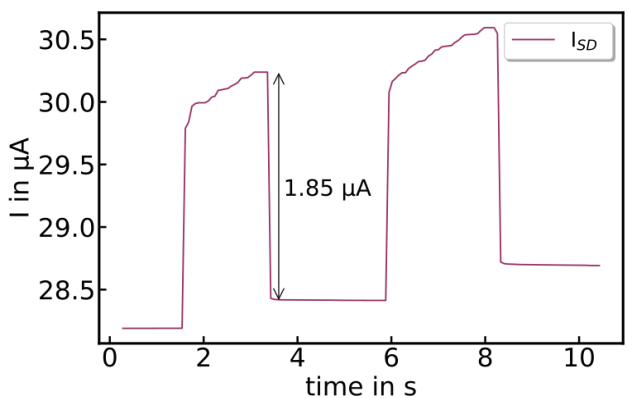

Figure 11: Photoresponse of a CsNb$_3$Br$_7$S crystal to a 779 nm laser at 200 K.

**Nuclear magnetic resonance**

The $^{133}$Cs MAS NMR spectrum exhibits a single peak at 330 ppm (Figure S3), near the upper edge of the chemical shift range for diamagnetic samples. The peak position is not far from that of CsBr (258 ppm),[38] consistent with the similar local Cs environment. The narrowness of the central peak indicates a small quadrupolar interaction, and the full extent of the spinning sideband envelope corresponds to a quadrupole coupling constant around 250 kHz, typical for crystalline materials.[39] The spinning sidebands at slower spinning rates are symmetrically distributed about the central peak, implying that there is no appreciable anisotropy in the Cs chemical shielding. No changes were observed in the central peak position or lineshape upon increasing the spinning rate, despite the accompanying temperature increase of about 25°C.[40] The most unique feature of the $^{133}$Cs NMR response is its anomalously short spin-lattice relaxation time: while most insulating and semiconducting solids have $^{133}$Cs $T_1$ values in the range of 30-60 s[41, 42] - with values up to 900 s having been reported [43] - the $T_1$ of CsNb$_3$Br$_7$S was measured to be 165 ms at 11.7 T. Such a short relaxation time is often indicative of nuclear interactions with unpaired electron spins, however, no evidence of a contact shift or electron-dipolar coupling is observed. Instead, this may be considered evidence for a Luttinger liquid, which exhibits extremely fast spin-lattice relaxation in certain magnetic field regimes.[44] A second $T_1$ measurement at 9.4 T yielded similarly rapid relaxation of 185 ms. Attempts to record a $^{93}$Nb NMR spectrum were partially successful, yielding a broad signal spanning nearly 1.8 MHz (not shown), reflective of the large quadrupole moment of $^{93}$Nb and its asymmetric local environments; however, the $^{93}$Nb $T_1$s are similarly short (< 200 ms) and the $T_2$ value estimated from the echo train is around 2 ms. These observations from NMR support the designation of CsNb$_3$Br$_7$S as a Luttinger liquid.

**Electronic Structure**

The electronic structure of CsNb$_3$Br$_7$S was examined further using density functional theory (DFT). The results are in qualitative agreement with earlier calculations (both extended Hückel[6] and DFT[45]) showing semimetallicity (specifically, zero band gap semiconductor behaviour) due to crossings of the Fermi level near the B (0 0 ½) and A (−½ 0 ½) points of the Brillouin zone (Figure 12Figure 13a). The band gap remains extremely small (< 0.1 eV) between B (0 0 ½) and D (0 ½ ½), A (−½ 0 ½), and E (−½ ½ ½).

In order to understand the nature of the electronic states near the Fermi level, we constructed maximally localized Wannier functions[46] from the DFT wavefunctions (Figure 12) in two ways. We first include all the bands with strong Nb d character (see fig. S5) in the energy window. This way we obtain Wannier functions that have a form like Nb d orbitals (Figure 12b), with additional of p-orbital tails coming from the hybridization with Br and S (open magenta circles in Figure 12). In this model, the largest absolute value for the hopping integral inside one trimer is around 1.2 eV, while the largest hopping for contiguous trimers can be as high as 0.7 eV, which underlines a non-negligible contribution of inter-trimer hybridization in the electronic structure.

The importance of considering contiguous trimers is more evident if we include only the 4 states around the Fermi energy in the energy window during the wannierization (full green circles in Figure 12). Figure 12 (c-d) shows the result of the spread minimization, with the Wannier functions being centred between the Nb trimers. The hopping integrals (t) of the tight-binding model corresponding to the bond-centred Wannier functions (Table S1) indicate that nearest-neighbour intra-chain

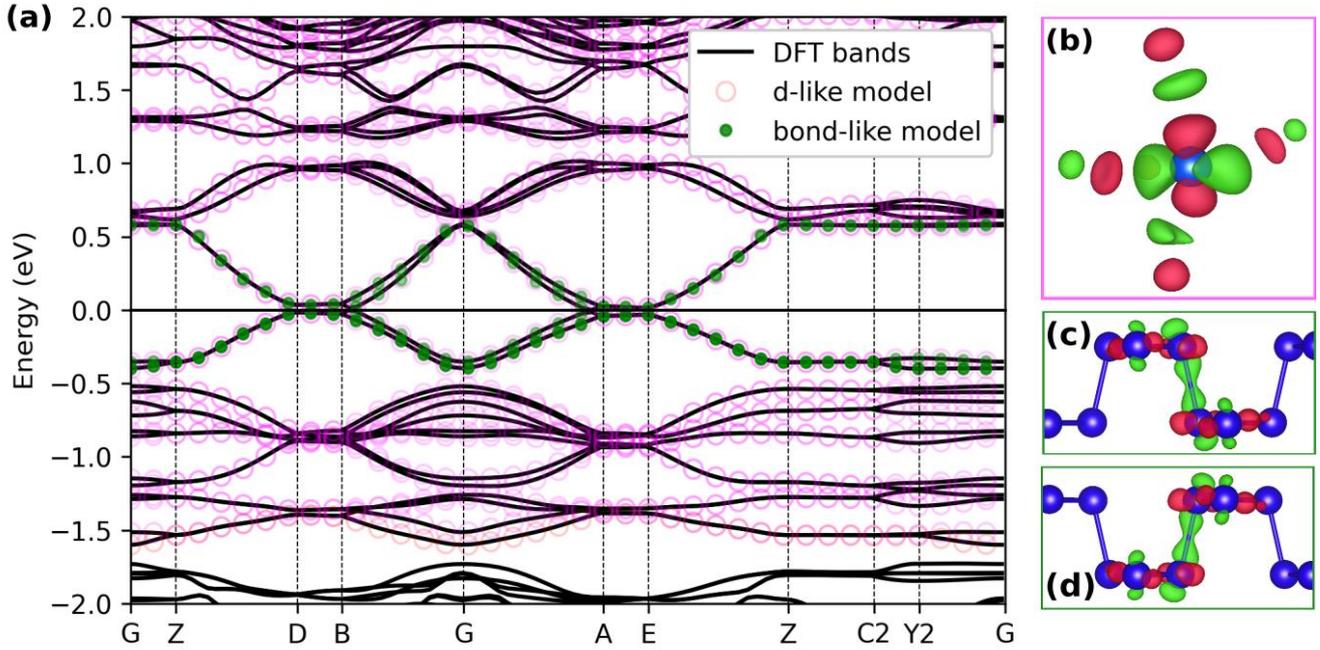

Figure 12: The calculated electronic band structure of nonmagnetic CsNb$_3$Br$_7$S (a), and corresponding atom-centred (b) and bond-centred (c–d) maximally localized Wannier functions (shown at Γ). Blue spheres represent Nb atoms. Special points in and paths through the Brillouin zone were chosen following literature.[36]

hopping (t ≈ −0.2 eV) is significantly favoured over inter-chain hopping (t' ≈ −0.01 eV).

The predicted electronic conduction in CsNb$_3$Br$_7$S is not unsurprising given the presence of unpaired electrons in each [Nb$_3$]$^{8+}$ cluster, which leads to each [Nb$_3$Br$_7$S]$^-$ chain having one unpaired electron per Nb$_3$ cluster. However, the measured temperature-activated electrical conductivity is inconsistent with a conventional metal or semimetal, but not with a Luttinger liquid. Together with the power law dependence of the electronic conductivity on temperature and short NMR T$_1$ relaxation times, our findings indicate that CsNb$_3$Br$_7$S is a Luttinger liquid above *ca.* 30 K. The values of the Luttinger parameter $\alpha$ extracted from the electronic conductivity ($\alpha$ = 4.4 for CsNb$_3$Br$_7$S and $\alpha$ = 4.2 for RbNb$_3$Br$_7$S) indicate that the electron–electron interactions are repulsive in nature, and that the chains contain strong impurities, *i.e.* discontinuities or barriers in the chains which require quantum tunnelling for electronic transport.[47] The presence of strong impurities precludes further modelling of the Luttinger electronic interaction parameters from the available data.

A metallic one-dimensional chain of electrons has several possible mechanisms of gap opening upon cooling. At low temperature, electronic correlations will favour the formation of a charge–ordered singlet (Peierls) or triplet (Mott) insulator, and which state forms depends on the relative energetics.[48] Below 30 K, as the conductivity no longer follows a power law, we expect a gap to have opened in the CDW and SDW continuum corresponding to the Luttinger liquid. We can therefore consider several mechanisms of gap opening in CsNb$_3$Br$_7$S.

The first is the Peierls distortion, which is a CDW instability that, for example, forms alternating long and short bonds (as in polyacetylene). We would expect this mechanism to lead to variations in the Nb–Nb bond lengths, either within the Nb$_3$ trimers as has been reported in Nb$_3$Cl$_8$[49] or between them.

In our structure refinement (at 100 K) each Nb$_3$ cluster in CsNb$_3$Br$_7$S is crystallographically identical; any Peierls distortion would lower the crystallographic symmetry and would be easily detectable using single-crystal X-ray diffraction. Therefore, we conclude that there is no Peierls distortion or charge disproportionation in CsNb$_3$Br$_7$S at 100 K, although there may be at lower temperatures.

Instead of forming bound pairs of electrons in a Peierls insulator, exchange coupling in combination with electronic correlations can create a Mott insulator *via* a SDW instability. Here, a Mott insulator would correspond to a state where each cluster has one localized electron with antiferromagnetic spin alignment with respect its neighbours along the chain (Figure 13). However, in a Mott insulator, the magnetic susceptibility would be expected to show temperature dependence corresponding to an antiferromagnet or a paramagnet with antiferromagnetic correlations, which is seen only below 30 K in CsNb$_3$Br$_7$S, and the material could be a Mott insulator below that temperature. Above 30 K, CsNb$_3$Br$_7$S shows temperature independent paramagnetic behaviour (TIP), and a magnetic-to-nonmagnetic charge transition seen in Nb$_3$Cl$_8$ is absent in CsNb$_3$Br$_7$S (Figure S4).

Spin-orbit coupling presents a third mechanism for opening a band gap. Previous studies have shown that, while in the absence of SOC, CsNb$_3$Br$_7$S is a topological nodal straight-line semimetal, inclusion of SOC introduces a small gap of *ca.* 5 meV at the crossing points. Note that due to the combination of P and T symmetries, the semimetal state in the absence of SOC has a well-defined, nontrivial $Z_2$ invariant,[50] and the gapped phase in the presence of SOC is also expected to be topological.[45] Such a gap is small, but still would be expected to lead to semiconducting behaviour for a Fermi liquid at low temperature. However, our calculations indicate that, while there is such a gap near B (0 0 ½), the bands cross the Fermi level near A (–½ 0 ½) (Figure 14). There is a small amount of dispersion of the electronic bands along $a$, which can be attributed to weak interchain interactions, and which leads to the formation of conducting electronic states.

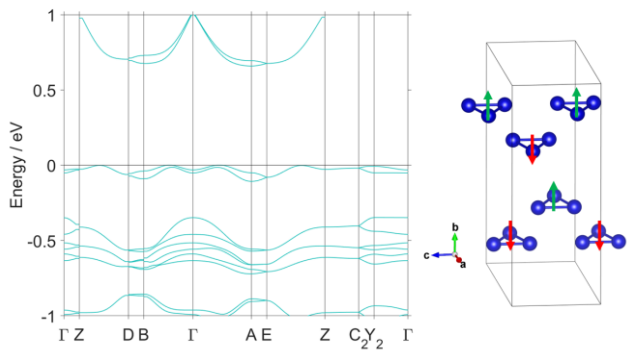

Figure 13: The calculated electronic band structure of $CsNb_3Br_7S$ with an antiferromagnetic arrangement of magnetic moments on the Nb triangles along each chain (shown at right). Special points in and paths through the Brillouin zone were chosen following literature.[36]

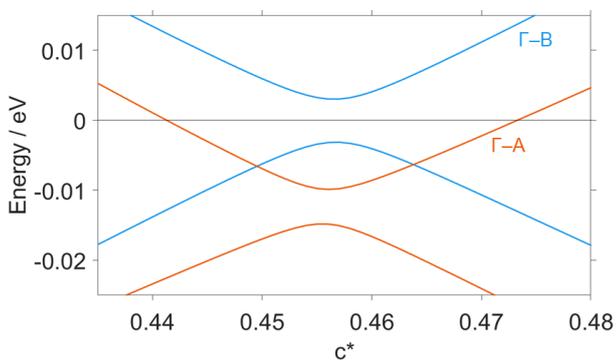

Figure 14: The calculated electronic band structure of nonmagnetic $CsNb_3Br_7S$, calculated with spin–orbit coupling on a fine grid of **k**-points along Γ–B (0 0 $c$*) (blue) and Γ–A (–½ 0 $c$*) (orange). Special points in and paths through the Brillouin zone were chosen following literature.[36]

Therefore, the nature of $CsNb_3Br_7S$ at temperatures below 30 K, where the Luttinger liquid state breaks down, cannot be definitively determined from the available data. The calculated phonon energies at Γ (Table S2) do not show an instability related to dimerization; the lowest energy phonon which involves modulation of the Nb–Nb distances has a frequency of 38.67 cm$^{-1}$ (4.6 meV). This indicates that the rough scale of the energy barrier which the CDW instability would need to overcome in order to create a Peierls distortion is small. The upturn in the magnetic susceptibility at low temperatures (Figure S4) suggests a potential Mott insulating state. Further low-temperature structural and magnetic investigation is therefore required to determine the ground states of $CsNb_3Br_7S$ and $RbNb_3Br_7S$ at temperatures approaching absolute zero.

## Conclusions

The previously given crystal structure of $CsNb_3Br_7S$ is confirmed by single crystal X-ray diffraction studies and the new compound $RbNb_3Br_7S$ is reported with a slightly distinct structure. Structure determinations on these compounds are challenging, due to the formation of twins and fraying character of needle-like rods parallel to the direction of infinite $[Nb_3Br_7S]^-$ chains. Crystals of both compounds appear black with metallic lustre but show semiconducting behaviour. Refinement of the crystal structure reveals a one-dimensional nature, due to the presence of $[Nb_3Br_7S]^-$ chains. However different arrangements of chains in the structure of $ANb_3Br_7S$ with A = Rb, Cs due to different ionic radii of A cations reflect significant ionic (A–Br) interaction between adjacent chains. This is also expressed by the calculated elastic properties (directional Young's modulus) of $CsNb_3Br_7S$ in which the chain direction ($c$-axis) is the stiffest direction (74 GPa), but local stiffness maxima also coincide with $a$ and $b$, indicating that ionic bonding between $[Nb_3Br_7S]^-$ chains through Cs cations is significant. Electronic structure calculations indicate that $CsNb_3Br_7S$ is a one-dimensional metal. At temperatures above 30 K, the experimental data indicates that the CDWs and SDWs are fluid, and the material behaves as a Luttinger liquid. The increase in conductivity with temperature in the Luttinger liquid phase indicates repulsive interactions between electrons. Below 30 K, a band gap could open through formation of a Peierls (CDW ordered) or Mott (SDW correlated) insulator.

## Conflicts of interest

There are no conflicts to declare.

## Acknowledgements

Support of this Research by the German Research Foundation (DFG) through grant ME 914/32-1 and SCHE1905/9-1 (project no. 426008387) is gratefully acknowledged. C.P.R. and S.F.W. were supported by ETH Zurich and by the European Union and Horizon 2020 through a Marie Sklodowska-Curie Fellowship, Grant Agreement No. 101030352 (C.P.R.) and by Grant No. 810451 (S. F. W.). A.C. was supported by ETH Zurich. Computational resources were provided by ETH Zurich, by the Swiss National Supercomputing Center (CSCS) under project IDs s1128 and eth3, and by the state of Baden-Württemberg through bwHPC and the DFG through grant INST 40/575-1 FUGG (JUSTUS 2 cluster). S.K. is supported by grants from the Natural Sciences and Engineering Research Council (NSERC) of Canada, and the Canada Foundation for Innovation (CFI). We thank Dr. Arun Krishnamurthy and Mr. Mojtaba Abbasi (University of Manitoba) for the solid-state NMR measurements.

## Notes and references

# Supporting Information

## Electronic Structure and Transport in the Potential Charge Transfer Insulators CsNb$_3$Br$_7$S and RbNb$_3$Br$_7$S

Fabian Grahlow[a], Fabian Strauß[b], Marcus Scheele[b], Markus Ströbele[a], Alberto Carta[c], Sophie F. Weber[c], Scott Kroeker[d], Carl P. Romao*[c] and H.-Jürgen Meyer*[a]

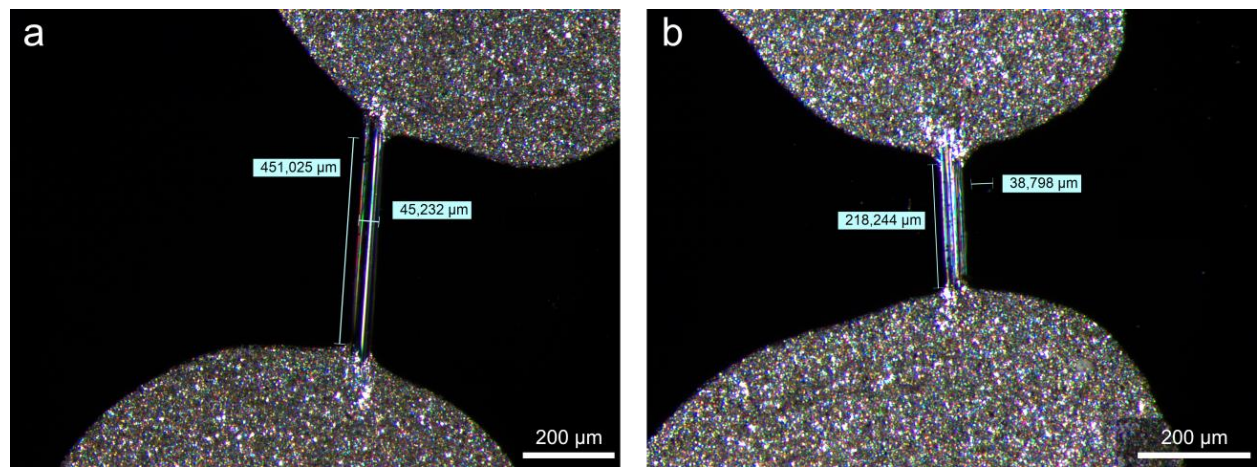

**Figure S1:** Optical micrographs of a (a) contacted CsNb$_3$Br$_7$S crystal and a (b) contacted RbNb$_3$Br$_7$S crystal.

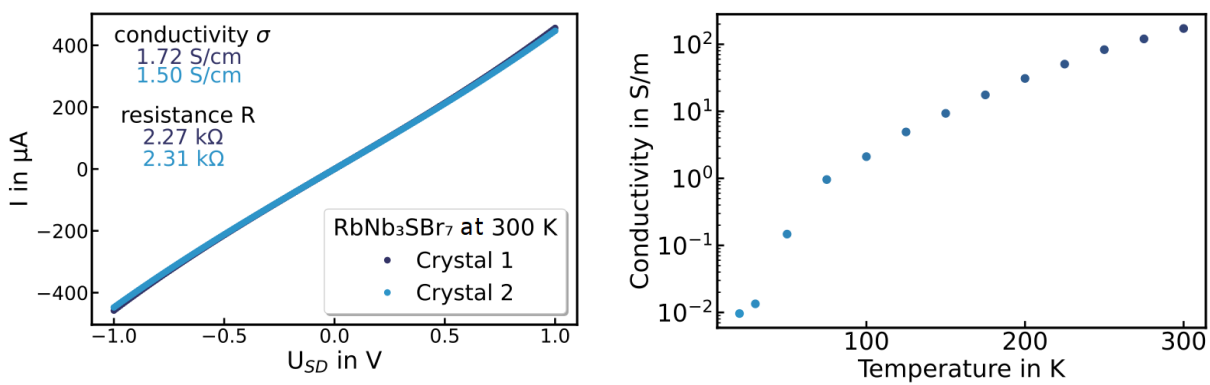

**Figure S2:** (left) Dark currents of RbNb$_3$Br$_7$S Crystals on silicon with 770 nm dioxide layer at 300 K. (Right) Electrical conductivity of RbNb$_3$Br$_7$S versus set temperature in a range of 20 K to 300 K.

[a.] Section for Solid State and Theoretical Inorganic Chemistry
Institute of Inorganic Chemistry
Eberhard-Karls-Universität Tübingen
Auf der Morgenstelle 18, 72076 Tübingen, Germany
*[a.] E-mail: juergen.meyer@uni-tuebingen.de
[b.] Institute for Physical and Theoretical Chemistry
Eberhard-Karls-Universität Tübingen
Auf der Morgenstelle 18, 72076 Tübingen, Germany
[c.] Department of Materials, ETH Zurich,
Wolfgang-Pauli-Str. 27, 8093 Zürich, Switzerland
*[c] E-mail: carl.romao@mat.ethz.ch
[d.] Department of Chemistry, University of Manitoba, Winnipeg, Manitoba R3T 2N2, Canada

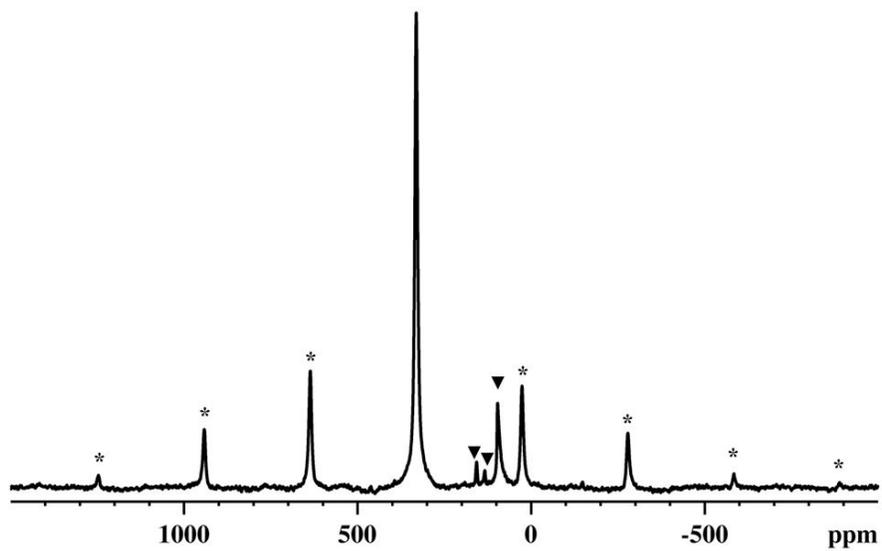

**Figure S3:** $^{133}$Cs MAS NMR spectrum of CsNb$_3$Br$_7$S acquired at 65.5 MHz with a spinning rate of 20.000(3) kHz. Spinning sidebands are marked with asterisks; peaks from minor impurities or decomposition products are indicated by arrows.

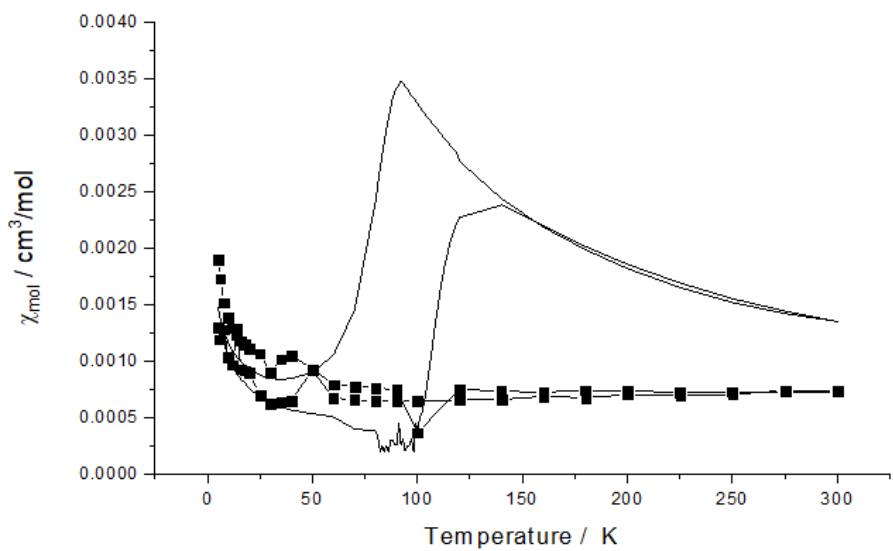

**Figure S4:** Measured magnetic susceptibility during one heating/cooling cycle at 100 Oe for Nb$_3$Cl$_8$ (line) and CsNb$_3$Br$_7$S (line with squares).

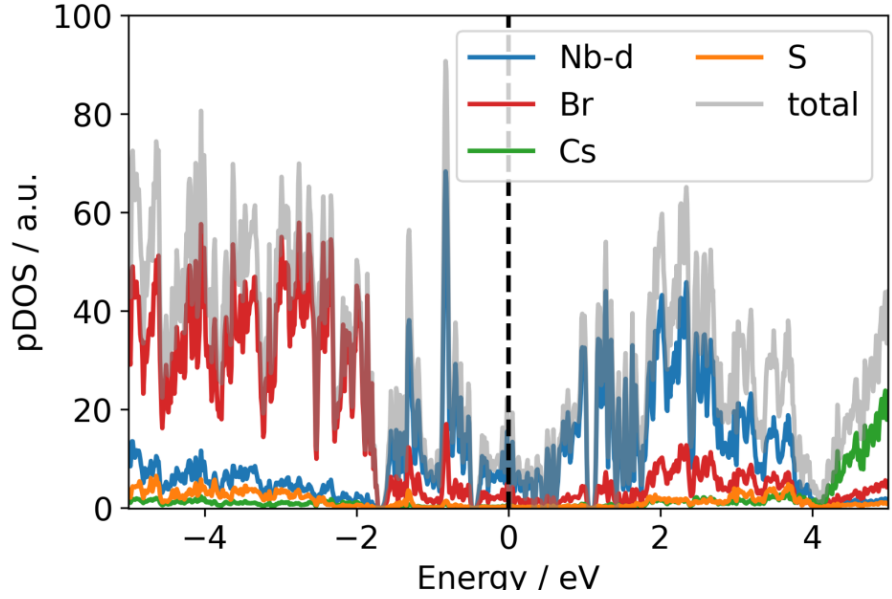

**Figure S5:** Calculated projected density of electronic states in CsNb$_3$Br$_7$S. The Fermi energy is shown as a black line.

**Table S1:** Parameters of the tight-binding model corresponding to the bond-centred Wannier functions of CsNb$_3$Br$_7$S.

| R (unit cell translation) | t (hopping integral) / eV | Note |
|---|---|---|
| [0 0 0] | 7.05 | On-site bond energy |
| [0 0 0] | -0.20 | Nearest neighbor intra-chain hopping |
| [0 0 1] | 0.02 | Next-nearest intra-chain hopping |
| [0 0 0] | -0.01 | Nearest inter-chain hopping |

**Table S2:** Calculated phonon frequencies at Γ of CsNb$_3$Br$_7$S.

| Phonon frequency / cm$^{-1}$ | | | | |
|---|---|---|---|---|
| 0.00 | 0.00 | 0.00 | 13.82 | 19.00 |
| 22.51 | 30.40 | 36.64 | 38.67 | 43.39 |
| 44.75 | 45.29 | 46.37 | 46.77 | 48.39 |
| 48.87 | 49.56 | 52.99 | 53.70 | 60.98 |
| 61.65 | 63.65 | 64.04 | 64.20 | 64.22 |
| 64.54 | 71.97 | 74.61 | 75.42 | 76.19 |
| 79.48 | 79.70 | 81.84 | 83.49 | 83.53 |
| 83.97 | 85.83 | 87.15 | 87.45 | 91.20 |
| 92.28 | 92.43 | 93.26 | 94.01 | 96.09 |
| 96.48 | 96.84 | 99.12 | 99.95 | 100.35 |
| 103.28 | 103.46 | 104.84 | 105.28 | 106.97 |
| 107.81 | 108.53 | 110.47 | 114.45 | 115.29 |
| 115.87 | 116.33 | 120.49 | 120.61 | 122.27 |
| 122.29 | 127.94 | 128.16 | 129.81 | 131.58 |
| 131.70 | 131.83 | 132.11 | 133.17 | 134.87 |
| 135.14 | 137.85 | 138.71 | 139.79 | 140.27 |
| 144.25 | 145.04 | 145.26 | 146.83 | 148.07 |
| 148.69 | 150.41 | 150.49 | 151.01 | 152.26 |
| 154.77 | 155.07 | 167.36 | 167.92 | 169.24 |
| 170.62 | 174.51 | 174.71 | 191.26 | 191.41 |
| 195.37 | 195.41 | 198.30 | 204.25 | 210.01 |
| 210.04 | 210.17 | 212.79 | 212.86 | 215.80 |
| 216.07 | 217.71 | 217.79 | 219.59 | 222.61 |
| 228.86 | 229.39 | 229.53 | 231.31 | 232.42 |
| 238.55 | 239.04 | 243.34 | 243.73 | 246.34 |
| 250.03 | 251.19 | 253.81 | 255.17 | 255.42 |
| 264.96 | 266.98 | 311.65 | 311.80 | 329.66 |
| 330.67 | 332.12 | 332.23 | 342.05 | 342.21 |
| 389.45 | 390.98 | 394.79 | 395.03 | |